\documentclass[prl,english,a4,superscriptaddress,twocolumn,amsmath,amssymb]{revtex4}
\usepackage{amsmath,amssymb,amsfonts,graphicx,multirow,color,bm}
\usepackage{bm,bbm}

\usepackage{hyperref,cleveref}
\usepackage{graphicx,color,relsize}

\def\beq{\begin{eqnarray}}
\def\eeq{\end{eqnarray}}
\def\beqa{\begin{eqnarray}}
\def\eeqa{\end{eqnarray}}

\newcommand{\cb}{\color{black}}

\newcommand{\clb}{\color{black}}

\begin{document}

\title{{\cb Mechanism for unconventional superconductivity in  the hole-doped \\ Rashba-Hubbard model}}

\author{Andr\'es Greco}
\email[]{agreco@fceia.unr.edu.ar}
\affiliation{Facultad de Ciencias Exactas, Ingenier\'{\i}a y Agrimensura and
Instituto de F\'{\i}sica Rosario (UNR-CONICET), Av. Pellegrini 250, 2000 Rosario, Argentina}
             
\author{Andreas P.\ Schnyder}
\email[]{a.schnyder@fkf.mpg.de}
\affiliation{Max-Planck-Institute for Solid State Research, D-70569 Stuttgart, Germany}

\date{\today}

\begin{abstract}
Motivated by the recent resurgence of interest in topological superconductivity, we study superconducting pairing 
instabilities of the hole-doped Rashba-Hubbard model on the square lattice with first- and second-neighbor hopping. 
Within the random phase approximation 
we compute the spin-fluctuation mediated paring interactions  
as a function of filling.
Rashba spin-orbit coupling splits the spin degeneracies of the bands, which leads to two van Hove singularities at two different fillings.
We find that for a  doping region in between these two van Hove fillings the spin fluctuations exhibit a strong ferromagnetic contribution. 
Because of these ferromagnetic fluctuations,  there is a strong tendency towards  spin-triplet $f$-wave pairing within this filling region, 
resulting in a topologically nontrival phase.
\end{abstract}

\maketitle

\vspace{-1.8cm}

Topological superconductors (TSCs) have attracted great interest recently
due to their potential use for quantum information technology and 
novel superconducting devices~\cite{chiu_RMP_16,hasan_rmp,qi_rmp,Schnyder2008gf,schnyder_brydon_review_JPCM_15,sato_ando_review_arXiv_16}. 
Many interesting topological phases, such as the  chiral $p$-wave state~\cite{read_green_PRB_00}, are realized in superconductors with odd-parity spin-triplet pairing.
However, until now only a few material systems have been discovered which show  spin-triplet superconductivity~\cite{kallin_berlinsky_rep_prog_phys_16,maeno_review_12,agterberg_review_17}, since spin-singlet pairing  is in most cases the dominant pairing channel. 
There are two types of TSCs with triplet pairing: intrinsic and artificial ones. 
While the former type arises as an intrinsic property of the material, the latter is artificially engineered in heterostructures by proximity coupling
to an $s$-wave superconductor~\cite{fu_kane_08}.
Intrinsic TSCs have the advantage that the topological phase exists in the entire volume of the material, and not just at an interface
of a heterostructure.
In recent years it has become clear that strong spin-orbit coupling (SOC) is conducive to triplet superconductivity~\cite{yanase_sigrist_JPSJ_07,yanase_sigrist_JPSJ_07b,yokoyama_tanaka_PRB_07}. Indeed, most candidate 
materials for intrinsic TSCs, such as  Sr$_2$RuO$_4$~\cite{kallin_berlinsky_rep_prog_phys_16,maeno_review_12}, CePt$_3$Si~\cite{agterberg_review_17},
and Cu$_x$Bi$_2$Se$_3$~\cite{hor_ong_cava_PRL,sasaki_ando_11}, contain heavy elements with strong spin-orbit interactions. 
Unfortunately, the strongly correlated TSCs Sr$_2$RuO$_4$ and CePt$_3$Si have a rather low $T_c$  of  $\lesssim 1~K$, 
while the pairing symmetry of the weakly correlated TSC Cu$_x$Bi$_2$Se$_3$ is still under debate~\cite{levy_stroscio_PRL_13,brydon_sau_PRB_14}. 
Therefore, the search for new intrinsic TSCs remains an important goal.

Parallel to these developments, MBE fabrication of oxide and heavy-fermion superlattices has seen great progress~\cite{chakhalian_science_07,reyren_mannhart_science_07,mizukami_matsuda_nat_phys_11,matsuda_PRL_14}. 
An important distinguishing feature of  epitaxial superlattices is their high tunability. That is, carrier density, Fermi surface (FS) topology,
as well as SOC can be tuned by modulating the layer thickness or by applying electric fields~\cite{caviglia_triscone_Nature_08,matsuda_PRL_14}. 
Remarkably, some of these superlattices show unconventional superconductivity with a fairly high-transition temperature.
One example is the heavy-fermion superlattice CeCoIn$_5$/YbCoIn$_5$~\cite{mizukami_matsuda_nat_phys_11,matsuda_PRL_14}, in which
 magnetic fluctuations~\cite{stock_magnetic_fluc_CeCoIn5} lead to 
superconductivity below $T_\textrm{c} \simeq 2$~K.  Modulating the layer thicknesses in this superlattice breaks the
inversion symmetry, which induces   Rashba spin-orbit interactions.
Interestingly, 
the strength of the Rashba SOC  can be controlled by the width of the  YbCoIn$_5$ block layers.
Strong Rashba interaction drastically alters the FS topology by splitting the spin degeneracy. This in turn is  favorable for
 triplet superconductivity, provided that the pairing mechanism allows for it. 
As is known from extensive theoretical works on cuprate superconductors~\cite{roemer_PRB_15,scalapino_hirsch_86,hlubina_PRB_99},
the shape and topology of the FS strongly influence the relative strengths of different pairing channels.
In order to optimize the layer thickness modulation in  CeCoIn$_5$/YbCoIn$_5$  for triplet superconductivity,
it is therefore important to understand the detailed interdependence among  Rashba SOC,
FS topology, and superconducting pairing symmetry.

\begin{figure}
\centering
\includegraphics[width=8.5cm,angle=0]{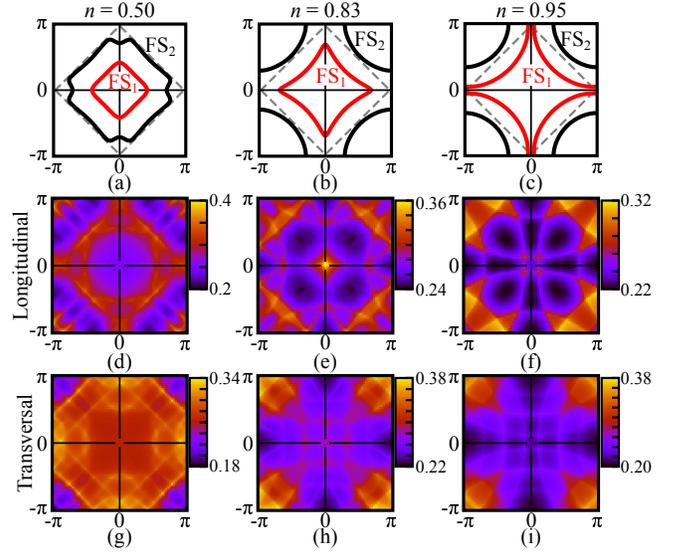}
\caption{ \label{mFig2}
 (a)-(c) {\cb Calculated} Fermi surface topology
 and (d)-(i) static $\omega=0$ spin susceptibility  
for the fillings $n=0.50$, $n=0.83$, and $n=0.95$, 
with $t^\prime = 0.3$, $V_{\textrm{so}}=0.5$,  $T = 0.01$, and $U=0.4$.
The second and third rows show the longitudinal and transversal susceptibilities, 
$\chi_{\textrm{long}}=\chi_{\uparrow \uparrow \uparrow \uparrow} - \chi_{\uparrow \downarrow \downarrow \uparrow}$ 
and 
$\chi_{\textrm{trans}} = \chi_{\uparrow \uparrow \downarrow \downarrow}$, 
respectively (see SM~\cite{supp_info}).  }
\label{fig:1}
\end{figure}

\begin{figure}
\centering
\includegraphics[width=9.cm,angle=-0]{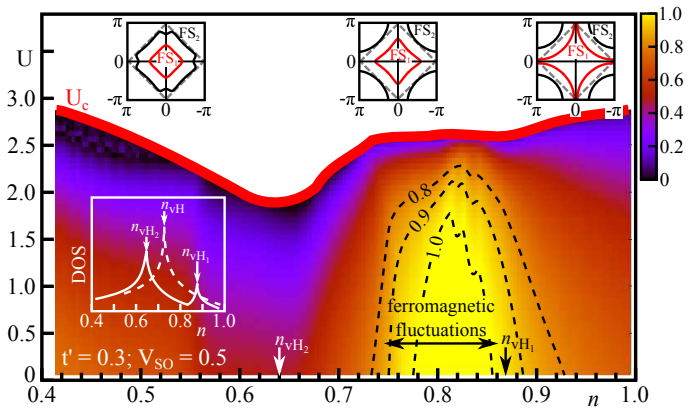}
\caption{ \label{mFig1}
The critical interaction strength $U_c$ as a function of
filling~$n$ is indicated by the red line.   
The color scale represents the relative intensity of the ferromagnetic fluctuations
in the longitudinal susceptibility.  
The inset shows the density of states versus filling for 
$V_{\textrm{so}}=0$ (dashed line) and $V_{\textrm{so}}=0.5$ (solid line). 
}
\label{fig:1}
\end{figure}

\begin{figure}
\centering
\includegraphics[width=9.cm,angle=0]{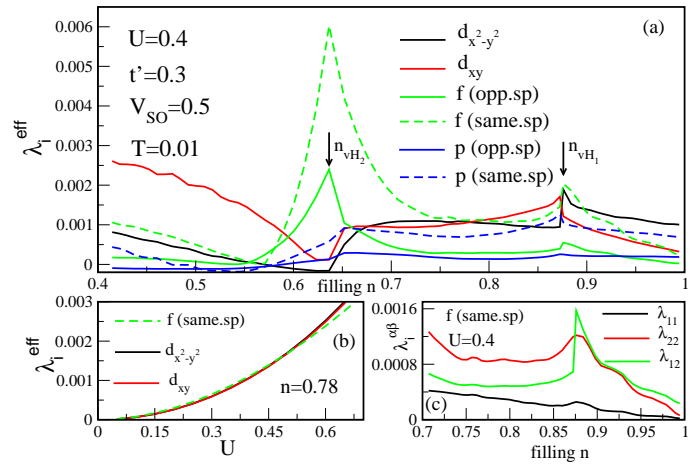}
\caption{ \label{mFig3}
(a) Filling dependence of the superconducting couplings $\lambda_i^{\textrm{eff}}$, 
as determined from Eq.~\eqref{pairing_strength} for $U=0.4$, and  
for the lowest-harmonic pairing symmetries given by Eq.~(S11). 
Here, we do not show the $s$-wave pairing
channel, since it is highly suppressed (i.e., negative) for the entire hole-doping range. 
For the numerical evaluation of Eq.~\eqref{pairing_strength} we used  $408$ Fermi momenta.
(b)  $\lambda^{\textrm{eff}}_i$ versus $U$  for the filling $n=0.78$. We present results
up to $U=0.7$ to show the regime where $f$-wave is dominant. With increasing $U$ the curves increase monotonically, and 
$d$-wave becomes dominant while $f$-wave subdominant for $U \gtrsim 0.5$. 
Near the magnetic instability $U = U_c$ we find that  
$\lambda_d^{\textrm{eff}} \sim 0.2$ and $\lambda_f^{\textrm{eff}} \sim 0.15$.
(c) Filling dependence of the {\clb intra and inter FS} pairing strengths, $\lambda^{\alpha \alpha}_i$ and $\lambda^{\alpha \beta}_i$,
for the $f$-wave channel with same spin projections.
}
\label{fig:2}
\end{figure}

Motivated by these considerations, we analyze in this Letter superconducting pairing instabilities of the hole-doped Rashba-Hubbard model,
which describes the essential features of many strongly correlated materials with Rashba 
SOC~\cite{yokoyama_tanaka_PRB_07,yanase_sigrist_JPSJ_07,yanase_sigrist_JPSJ_07b,riera_PRB_14}.
Focusing on the square lattice with first- and second-neighbor hopping, $t$ and $t^{\prime}$, we 
compute
the spin-fluctuation-mediated pairing interaction  as a function of filling. 
{\clb For this purpose we use the random phase approximation (RPA), which is known to qualitatively capture the essential physics, at least within weak 
coupling~\cite{yanase_sigrist_JPSJ_07,yanase_sigrist_JPSJ_07b,yokoyama_tanaka_PRB_07,roemer_PRB_15,scalapino_hirsch_86,hlubina_PRB_99}.}
Finite SOC splits the energy bands leading to two van Hove singularities at the fillings $n_{vH_1}$ and $n_{vH_2}$.
Remarkably, we find that in a  doping region in between these two van Hove fillings, there exist strong 
\emph{ferromagnetic} (FM) spin fluctuations (Figs.~\ref{mFig2} and~\ref{mFig1}).
Due to these FM fluctuations   there is a strong tendency towards  spin-triplet $f$-wave pairing 
in this filling region, while
the pairing channels of $d$-wave type (Fig.~\ref{mFig3}) are of the same order or subdominant.

 \textit{Model and Method.---} 
The Rashba-Hubbard model on the square lattice is given  
\begin{equation} 
H = \sum_{\bf k} {\psi}^\dag_{\bf k}  \hat{h} ( {\bf k} )    \psi^{\ }_{{\bf k}}  
+  U \sum_{{\bf k},{\bf k'}, {\bf q}} c^\dag_{{\bf k} \uparrow} c_{{\bf k}+{\bf q} \uparrow}
c^\dag_{{\bf k'} \downarrow}c_{{\bf k'}-{\bf q} \downarrow}, 
\label{rashba_hubbard}
\end{equation}
where $U$ is the local Coulomb repulsion,
$\hat{h}({\bf k} ) = \left( \varepsilon_{\bf k} \tau_0 + {\bf g}_{\bf k} \cdot \boldsymbol{\tau}   \right)$, 
and $\psi_{\bf k}=(c_{{\bf k} \uparrow}, c_{{\bf k} \downarrow})^T$.
Here, 
$\boldsymbol{\tau} = (\tau_1, \tau_2, \tau_3)^T$ are the three Pauli matrices, 
and $\tau_0$ stands for the $2\times2$ unit matrix.
The band energy $\varepsilon_{\bf k}=-2t(\cos k_x +\cos k_y) + t' \cos k_x  \cos k_y -\mu$
contains both first- and second-neighbor hopping, $t$ and $t'$, respectively, and
is measured relative to the chemical potential $\mu$.
The vector ${\bf g}_{\bf k}$ describes Rashba SOC with
 ${\bf g}_{\bf k}=V_{\textrm{so}}(\partial \varepsilon_{\bf k} / \partial k_y,  - \partial \varepsilon_{\bf k} / \partial k_x ,0)$
and the coupling constant $V_{\textrm{so}}$.
For our numerical calculations we set  $t=1$, $t'=0.3$, and $V_{\textrm{so}}=0.5$,
and focus on the hole-doped case with filling
$0.4 < n<1$.
We have checked that other parameter choices do no qualitatively change our findings.
The presence of Rashba SOC splits
the electronic dispersion $\varepsilon_{\bf k}$ into negative- and positive-helicity bands
with energies 
$E_{\bf k}^{1} = \varepsilon_{\bf k}  - \left| {\bf g}_{\bf k} \right|$ and $E_{\bf k}^{2} = \varepsilon_{\bf k}  + \left| {\bf g}_{\bf k} \right|$, 
respectively.
Both spin-split bands exhibit   van Hove singularities at ${\bf k} = (\pi,0)$ and symmetry related points. 
For our parameter choice the corresponding van Hove fillings occur at $n_{vH_1} \simeq 0.87$ and
$n_{vH_2} \simeq 0.65$, see inset of Fig.~\ref{mFig1}. 

The first term of Eq.~\ref{rashba_hubbard} defines the bare $2 \times 2$ fermionic Greens function
in the spin basis
\begin{equation}
G^{(0)}_{\sigma_1 \sigma_2} ( {\bf k} , i \nu_n ) = \left(\left[ i \nu_n  \tau_0 - \hat{h} ({\bf k} ) \right]^{-1} \right)_{\sigma_1 \sigma_2} ,
\end{equation}
where $\nu_n=  2 n \pi / \beta$ is the fermionic Matsubara frequency.  
For $U=0$ the bare spin susceptibility can be expressed in terms of $G^{(0)}$ as
\begin{equation}
\chi^{(0)}_{\sigma_1 \sigma_2 \sigma_3 \sigma_4} ({\bf q},i\omega_l)=
\sum_{ {\bf k}, i\nu_n} G^{(0)}_{\sigma_1 \sigma_2}  ({\bf k},i\nu_n) 
G^{(0)}_{\sigma_3 \sigma_4}  ( {\bf k}+ {\bf q},i\nu_n+i\omega_l) ,
\end{equation}
\noindent where $\omega_l=2 l\pi/\beta$ is the bosonic Matsubara frequency.
Within the RPA ~\cite{yanase_sigrist_JPSJ_07,yanase_sigrist_JPSJ_07b}
the dressed spin susceptibility $\chi_{\sigma_1 \sigma_2 \sigma_3 \sigma_4} ({\bf q},i\omega_l)$ 
is computed as
\begin{equation} \label{eq_RPA_suscep}
\hat{\chi} ({\bf q},i\omega_l)
=
\left[ \mathbbm{1}  -   \hat{\chi}^{(0)} ( {\bf q}, i \omega_l )   \hat{U} \right]^{-1} 
\hat{\chi}^{(0)} ({\bf q},i\omega_l) .
\end{equation}
In Eq.~\eqref{eq_RPA_suscep} the sixteen components of $\chi_{\sigma_1 \sigma_2 \sigma_3 \sigma_4}$ and $\chi^{(0)}_{\sigma_1 \sigma_2 \sigma_3 \sigma_4}$
are stored in the $4 \times 4$ matrices  $\hat{\chi}$ and $\hat{\chi^{(0)}}$, respectively, and the 
$4 \times 4$ coupling matrix 
$\hat{U}$ is antidiagonal, see the Supplemental Material (SM) \cite{supp_info} for details.

The spin fluctuations described by Eq.~\eqref{eq_RPA_suscep} can lead to an effective interaction
that combines two electrons into a Cooper pair. As in Refs.~\onlinecite{roemer_PRB_15,scalapino_hirsch_86}, it is necessary 
to distinguish between the interaction for same  [$V_{\textrm{same}}^{\textrm{eff}}({\bf k},{\bf k'})$] and for 
opposite [$V_{\textrm{opp}}^{\textrm{eff}}({\bf k},{\bf k'})$]
spin projections 
between two electrons  with momentum ${\bf k}$ and ${\bf k'}$\cite{footnoteWeakCoupling}, 
which are given in the spin basis by
\begin{subequations}
\begin{eqnarray}
V^{\textrm{eff}}_{\textrm{same}}({\bf k}, {\bf k'})
&=&
U^2 \chi_{\sigma \sigma \sigma \sigma} ({\bf k-k'}) ,
\\
V^{\textrm{eff}}_{\textrm{opp}}({\bf k}, {\bf k'})
&=&
U^2 \chi_{\sigma \bar{\sigma} \bar{\sigma} \sigma} ({\bf k-k'})+
U^2 \chi_{\sigma \sigma \bar{\sigma} \bar{\sigma }} ({\bf k+k'}) , \quad
\end{eqnarray}
\end{subequations}
respectively.
In weak coupling we can define for each pairing channel $i$
a dimensionless pairing strength  as~\cite{frigeri_sigrist_EPJB_2006,dolgov_golubov_PRB_08,dolgov_mazin_PRB_09,hirschfeld_mazin_rep_prog_11}
\begin{equation} \label{pairing_strength}
\lambda_i^{\alpha \beta}
= 
- \frac{  
\mathlarger{\int}_{\textrm{FS}_\alpha} 
\frac{ d k}{ v^\alpha_{\textrm{F}} ( k ) }  
\mathlarger{\int}_{\textrm{FS}_\beta} 
\frac{ d k'}{ v^\beta_{\textrm{F}} ( k' ) } 
\eta_i (k) 
 V_{\textrm{s}/\textrm{t}}^{\textrm{eff}} (k, k')
\eta_i (k')
}
{
  2 \pi^2
\mathlarger{\int}_{\textrm{FS}_\beta} 
\frac{ d k^{\prime} }{ v^\beta_{F} (k^{\prime} ) }  
\left[ \eta_i(k^{\prime} ) \right]^2
}, 
\end{equation}
{\clb where  $\alpha$ and $\beta$ label the FS sheets}. 
The diagonal and off-diagonal elements of $\lambda_i^{\alpha \beta}$ represent
{\clb intra- and inter Fermi surface pairing strengths,} respectively. 
In Eq.~\eqref{pairing_strength},  $k$ and $k'$ are restricted to the Fermi sheets  $\textrm{FS}_\alpha$ and 
 $\textrm{FS}_\beta$, respectively,     $v^{\alpha}_{\textrm{F}} ( {\bf k} )  = \left| \nabla_{\bf k} E^{\alpha}_{\bf k} \right|$
is the Fermi velocity, 
and $\eta_i (k)$ describes the
$k$ dependence of each possible pairing symmetry, see SM~\cite{supp_info}.
In the case of singlet pairing the effective interaction in Eq.~\eqref{pairing_strength}
is solely due to scattering between electrons with opposite spins.
For triplet pairing, however, both same- and opposite-spin scattering processes 
are possible.
The effective superconducting coupling constant $\lambda_i^{\textrm{eff}}$ for a given 
pairing channel $i$ is given by the largest  eigenvalue
of the matrix $\lambda_i^{\alpha \beta}$~\cite{dolgov_golubov_PRB_08}.
Hence, by numerically evaluating Eq.~\eqref{pairing_strength} for all possible channels 
$i$ we can identify the leading pairing instability
as a function of filling  and SOC strength.

 %%%%%%%%%%%%%%%%%%%%%%%%%%%%%%
%%%%%%%%%%%%%%%%%%%%%%%%%%%%%%

\textit{Spin susceptibility.---}
Before discussing superconductivity, let us first consider the static susceptibility in the paramagnetic state for intermediate 
coupling $U=0.4$ and $T=0.01$. While Figs.~\ref{mFig2}(d)-\ref{mFig2}(f) show the longitudinal susceptibility, 
Figs.~\ref{mFig2}(g)-\ref{mFig2}(i) show the transversal susceptibility for the fillings 
$n=0.5$, $n=0.83$, and $n=0.95$, respectively. As expected, and different to the case without SOC, the longitudinal 
and transversal susceptibilities show different spin texture. The FS topology for each filling 
is shown in Figs.~\ref{mFig2}(a)-\ref{mFig2}(c). 
The spin susceptibility shows  large magnetic fluctuations, whose magnetic modulation vectors ${\bf q}$ 
depend strongly on filling $n$ and
FS topology. Indeed, 
we observe an intricate interplay between FS topology and the structure of the spin susceptibility:
For $n > n_{vH_1}$ the two spin-split FS sheets are hole-like and centered at $(\pi,\pi)$ [Fig.~\ref{mFig2}(c)], 
which results
in  a spin susceptibility with incommensurate anti-ferromagnetic 
modulation vector  ${\bf q}=(\pi,\pi \pm \delta)$, see Figs.~\ref{mFig2}(f), and~\ref{mFig2}(i).
For $n <n_{vH_2}$, on the other hand,  both FS
sheets are electron-like and centered at   $\Gamma$ [Fig.~\ref{mFig2}(a)] leading
to a longitudinal spin susceptibility with nearly commensurate anti-ferromagnetic ${\bf q}$ vector  
[Fig.~\ref{mFig2}(d)]. In between the two van Hove fillings, $n_{vH_2} < n < n_{vH_1}$,
FS$_1$  is electron-like and centered at $\Gamma$, while FS$_2$
is hole-like and centered at $(\pi, \pi)$, see Fig.~\ref{mFig2}(b).
Interestingly, within this filling range there exists a 
broad region, i.e., $0.76 \lesssim n \lesssim n_{vH_1}$, where the dominant longitudinal fluctuations 
are ferromagnetic with ${\bf q}=(0,0)$, see Figs.~\ref{mFig2}(e) and~\ref{mFig1}.

Increasing the Hubbard interaction $U$ enhances the magnetic fluctuations and eventually drives 
the system into the magnetically ordered phase. 
In this process the modulation vector of the strongest fluctuations becomes the ordering wave vector of the
ordered phase. The transition between paramagnetic and ordered phase occurs at the critical interaction strength $U_c$
with a given ordering wave vector where the susceptibility diverges.
 Although the transversal and longitudinal susceptibilities
show different spin texture, both diverge simultaneously at the same ordering momentum, showing the 
non trivial feedback between them
for finite SOC. 
Figure~\ref{mFig1} displays the filling dependence of the critical interaction $U_c$ (red line).
The color scale indicates the intensity of the ferromagnetic fluctuations in the longitudinal susceptibility relative 
to the (incommensurate) antiferromagnetic fluctuations. We observe that the ferromagnetic fluctuations 
are dominant in the filling range $0.76 \lesssim n \lesssim n_{vH_1}$
and for $U$ within the range $0 \leq U \lesssim 1.6$. 
These ferromagnetic fluctuations originate from the combined effect of   finite SOC $V_{\textrm{so}}$ and
finite $t^{\prime}$. As a matter of fact, for  $V_{\textrm{so}}=0$ and $t^{\prime} \ne 0$ 
there is only one van Hove filling at  $n_{vH} \sim 0.72$ 
(inset of Fig.~\ref{mFig1}), 
which separates commensurate  antiferromagnetism [${\bf q}=(\pi,\pi)$] for $n > n_{vH}$  from incommensurate 
antiferromagnetism  [${\bf q}=(\pi,\pi-\delta)$] for $n < n_{vH}$, and 
ferromagnetic fluctuations only occur   in a narrow region around the van Hove filling 
$n_{vH}$~\cite{roemer_PRB_15}.
For  $t^{\prime} = 0$ and $V_{\textrm{so}} \ne 0$ ferromagnetic fluctuations are 
absent~\cite{yokoyama_tanaka_PRB_07}.
Different to the longitudinal susceptibility the transversal susceptibility shows 
ferromagnetic fluctuations only very close to the van Hove fillings $n_{vH_1}$ and $n_{vH_2}$ 
[Fig.~S5 of the SM].

%%%%%%%%%%%%%%%%%%%%%%%%%%%%%%
%%%%%%%%%%%%%%%%%%%%%%%%%%%%%%

\textit{Superconducting instabilities.---}
The discussed magnetic fluctuations can lead to superconducting pairing instabilities.
We set the 
Hubbard interaction to $U=0.4 < U_c$ and compute
$\lambda_i^{\textrm{eff}}$ within the filling range $0.4 <n < 1$ 
for the lowest-harmonic pairing symmetries, 
as defined in Eq.~(S11) of the SM. 
The resulting filling dependence of the pairing symmetries is presented in Fig.~\ref{mFig3}.
Note that the weak coupling approach is more reliable away from the van Hove fillings.
 At the van Hove fillings $n_{vH_i}$, $\lambda_i^{\textrm{eff}}$ exhibits large jumps 
due to the divergent density of states~\cite{footnote_RPA_at_van_Hove}, {\clb which is an artifact 
of the weak coupling approach.}
Let us examine the results of Fig.~\ref{mFig3} separately for (i) $n>n_{vH_1}$, 
(ii) $n < n_{vH_2}$, and (iii) $n_{vH_2}  < n < n_{vH_1}$:

(i) $n>n_{vH_1}$:
For this filling region the singlet d$_{x^2-y^2}$-wave pairing channel is dominant.
This is due to large anti-ferromagnetic spin fluctuations that exist in the entire hole-doping range $1>n>n_{vH_1}$,
similar to the case of $V_{\textrm{so}} =0$~\cite{roemer_PRB_15}.
The subleading pairing solutions have $p$-wave 
and $f$-wave symmetry due to effective interactions with same spin projections. Notice that
{\clb in contrast} to the case for $V_{\textrm{so}} =0$~\cite{roemer_PRB_15}, here the pairing strength for 
same spin projections is different from opposite spin projections.  
While the tendency  to superconductivity in the $f$-wave channel is 
strongly decreasing approaching half-filling, it is rather stable for the $p$-wave channel. 
Because Rashba SOC breaks inversion symmetry, we expect that in this filling range
the pairing symmetry is an admixture of $d_{x^2-y^2}$-wave, $p$-wave, and $f$-wave~\cite{frigeri_sigrist_PRL_04}.
However, since $\lambda^{\textrm{eff}}_{d_{x^2-y^2}} \gg \lambda^{\textrm{eff}}_{p/f}$, 
the $d_{x^2-y^2}$-wave channel is the leading one.

(ii) $n < n_{vH_2}$:
In this filling region the $d_{xy}$-wave pairing is leading, while the $f$-wave and $p$-wave channels are subdominant.
We ascribe this   tendency towards $d_{xy}$-wave pairing, rather than $d_{x^2-y^2}$-wave, to
the strong transversal spin fluctuations, which are peaked at $(\pi, 0)$ and symmetry related points.

(iii) $n_{vH_2} <  n < n_{vH_1}$: This is the most interesting region.
Remarkably, we find that around the filling  $n \simeq 0.78$
the triplet $f$-wave solution for same spin projections is the leading one, which we attribute
to the strong ferromagnetic fluctuations that occur for this filling 
in the longitudinal susceptibility [cf.~Figs.~\ref{mFig2}(e) and~\ref{mFig1}].
The subdominant pairing channels have $d_{x^2-y^2}$-wave and $d_{xy}$-wave form.
Hence, due to broken inversion symmetry, the gap is expected to exhibit also $d$-wave
admixture to the dominant $f$-wave harmonic. {\clb Although the weak coupling RPA approach
underestimates the values of $\lambda_i^{\textrm{eff}}$, it nevertheless  
qualitatively captures the relative tendency to superconductivity in each channel.}
Different to the case for $V_{\textrm{so}} = 0$ ~\cite{roemer_PRB_15}, where ferromagnetic fluctuations
occur only very close to the van Hove filling and a singular behavior is found at this filling 
for triplet $f$-wave pairing, here triplet $f$-wave extends in a broad filling region.  
This fact rules out the possibility that the observed tendency towards $f$-wave pairing is an artifact of the van Hove singularity. 
Without second-neighbor hopping the triplet pairing component is always subdominant~\cite{yokoyama_tanaka_PRB_07}.
Thus, our results offer a microscopic mechanism for the realization of triplet pairing with same spin projection, which was 
proposed on phenomenological grounds to be a candidate in non-centrosymmetric 
systems with strong SOC \cite{frigeri_sigrist_PRL_04}.

To analyze the dominance of the triplet $f$-wave channel
we show in Fig.~\ref{mFig3}(b) the dependence of $\lambda^{\textrm{eff}}_i$
on the interaction strength $U$ for  $n=0.78$.
We find that $\lambda^{\textrm{eff}}_f$ is the largest effective coupling
for $0.0 < U \lesssim 0.5$. This behavior is consistent with the result of Fig.~\ref{mFig1}
which shows that the ferromagnetic fluctuations become less and less dominant with increasing $U$.  
Before concluding, let us briefly discuss the contributions
of the {\clb intra and inter FS} scattering processes to the effective superconducting coupling.
In Figs.~\ref{mFig3}(c) we present the filling dependence of
the {\clb intra FS} ($\lambda^{11}_i$ and $\lambda^{22}_i$) and the {\clb inter FS} ($\lambda^{12}_i$)
pairing strengths for the $f$-wave channel for same spin projections~\cite{supp_info}. 
We observe that the $f$-wave pairing is driven by
{\clb intra FS} processes within FS$_2$. 

%%%%%%%%%%%%%%%%%%%%%%%%%%%%%%
%%%%%%%%%%%%%%%%%%%%%%%%%%%%%%

{\it Conclusions and implications for experiments.---}
We have studied superconducting instabilities of the hole-doped  Rashba-Hubbard model with first- and second-neighbor hopping
within a spin-fluctuation-mediated pairing scenario.
Using an RPA approach, we have determined the pairing symmetry as a function of filling and have shown
that there exists an interplay between   FS topology,   structure of the magnetic fluctuations, and  pairing symmetry.
In between the two van Hove fillings, close to $n \simeq 0.78$, 
the leading pairing solutions has 
 triplet $f$-wave symmetry, which is driven by ferromagnetic fluctuations. 
Since  within the spin fluctuation scenario the pairing symmetry is largely determined by the type of spin fluctuations,
we expect that more sophisticated treatments, such as FLEX~\cite{YANASE20031} or fRG~\cite{fRG_review_metzner_RMP_12}, will confirm our RPA analysis. 
The tendency towards $f$-wave pairing near  $n \simeq 0.78$ unavoidably leads to a topologically nontrivial  state.
The precise nature of this topological state depends on the  
detailed momentum structure of the gap.
There are three possibilities: (i) The superconducting state is nodal with a dominant $f$-wave pairing symmetry and only small admixtures of $d$-wave and $p$-wave components. 
The point nodes of this superconducting state are topologically protected by a winding number, which gives rise to Majorana flat band edge states~\cite{schnyder_brydon_review_JPCM_15}. 
(ii) The superconducting state is fully gapped due to a sizable admixture of $d$-wave and $p$-wave components. In this case the superconducting state belongs to symmetry class DIII and
exhibits helical Majorana edge states~\cite{chiu_RMP_16}.
(iii) The non-linear gap equation has a complex solution, yielding a time-reversal breaking triplet pairing state without nodes. This corresponds to a topological superconductor in symmetry class D, with 
chiral Majorana edge states~\cite{chiu_RMP_16}.
{\cb In closing we note that pair decoherence caused by impurity scattering is suppressed in all of the above three scenarios, due
to  the spin-momentum locking of the band structure~\cite{footnote_disorder}.}

Our findings provide a new mechanism for the creation of triplet superconductivity, which is relevant for non-centrosymmetric 
superconductors with strong SOC~\cite{agterberg_review_17,schnyder_brydon_review_JPCM_15} and for oxide and heavy-fermion hybrid structures~\cite{chakhalian_science_07,reyren_mannhart_science_07,mizukami_matsuda_nat_phys_11,matsuda_PRL_14}. 
It might be possible to realize the discussed $f$-wave state in CeCoIn$_5$/YbCoIn$_5$
hybrid structures~\cite{mizukami_matsuda_nat_phys_11,matsuda_PRL_14},
by an appropriate choice
of layer thickness modulation.
We hope that the present study will stimulate further experimental investigations along these directions.

 \emph{Acknowledgments.---}
We thank M.~Bejas, H.~Nakamura,  J.~Riera, A.~T.~ R{\o}mer, T.~Yokoyama, 
and R.~Zeyher for useful discussions. 
{\cb A.G.~thanks the Max Planck Institute for Solid State Research in Stuttgart for hospitality and financial support.}

\bibliographystyle{apsrev}
\bibliography{rashba-hubbard}

\begin{thebibliography}{40}
\expandafter\ifx\csname natexlab\endcsname\relax\def\natexlab#1{#1}\fi
\expandafter\ifx\csname bibnamefont\endcsname\relax
  \def\bibnamefont#1{#1}\fi
\expandafter\ifx\csname bibfnamefont\endcsname\relax
  \def\bibfnamefont#1{#1}\fi
\expandafter\ifx\csname citenamefont\endcsname\relax
  \def\citenamefont#1{#1}\fi
\expandafter\ifx\csname url\endcsname\relax
  \def\url#1{\texttt{#1}}\fi
\expandafter\ifx\csname urlprefix\endcsname\relax\def\urlprefix{URL }\fi
\providecommand{\bibinfo}[2]{#2}
\providecommand{\eprint}[2][]{\url{#2}}

\bibitem[{\citenamefont{Chiu et~al.}(2016)\citenamefont{Chiu, Teo, Schnyder,
  and Ryu}}]{chiu_RMP_16}
\bibinfo{author}{\bibfnamefont{C.-K.} \bibnamefont{Chiu}},
  \bibinfo{author}{\bibfnamefont{J.~C.~Y.} \bibnamefont{Teo}},
  \bibinfo{author}{\bibfnamefont{A.~P.} \bibnamefont{Schnyder}},
  \bibnamefont{and} \bibinfo{author}{\bibfnamefont{S.}~\bibnamefont{Ryu}},
  \bibinfo{journal}{Rev. Mod. Phys.} \textbf{\bibinfo{volume}{88}},
  \bibinfo{pages}{035005} (\bibinfo{year}{2016}),
  \urlprefix\url{https://link.aps.org/doi/10.1103/RevModPhys.88.035005}.

\bibitem[{\citenamefont{Hasan and Kane}(2010)}]{hasan_rmp}
\bibinfo{author}{\bibfnamefont{M.~Z.} \bibnamefont{Hasan}} \bibnamefont{and}
  \bibinfo{author}{\bibfnamefont{C.~L.} \bibnamefont{Kane}},
  \bibinfo{journal}{Rev. Mod. Phys.} \textbf{\bibinfo{volume}{82}},
  \bibinfo{pages}{3045} (\bibinfo{year}{2010}).

\bibitem[{\citenamefont{Qi and Zhang}(2011)}]{qi_rmp}
\bibinfo{author}{\bibfnamefont{X.-L.} \bibnamefont{Qi}} \bibnamefont{and}
  \bibinfo{author}{\bibfnamefont{S.-C.} \bibnamefont{Zhang}},
  \bibinfo{journal}{Rev. Mod. Phys.} \textbf{\bibinfo{volume}{83}},
  \bibinfo{pages}{1057} (\bibinfo{year}{2011}).

\bibitem[{\citenamefont{Schnyder et~al.}(2008)\citenamefont{Schnyder, Ryu,
  Furusaki, and Ludwig}}]{Schnyder2008gf}
\bibinfo{author}{\bibfnamefont{A.~P.} \bibnamefont{Schnyder}},
  \bibinfo{author}{\bibfnamefont{S.}~\bibnamefont{Ryu}},
  \bibinfo{author}{\bibfnamefont{A.}~\bibnamefont{Furusaki}}, \bibnamefont{and}
  \bibinfo{author}{\bibfnamefont{A.~W.~W.} \bibnamefont{Ludwig}},
  \bibinfo{journal}{Phys. Rev. B} \textbf{\bibinfo{volume}{78}},
  \bibinfo{pages}{195125} (\bibinfo{year}{2008}).

\bibitem[{\citenamefont{Schnyder and
  Brydon}(2015)}]{schnyder_brydon_review_JPCM_15}
\bibinfo{author}{\bibfnamefont{A.~P.} \bibnamefont{Schnyder}} \bibnamefont{and}
  \bibinfo{author}{\bibfnamefont{P.~M.~R.} \bibnamefont{Brydon}},
  \bibinfo{journal}{Journal of Physics: Condensed Matter}
  \textbf{\bibinfo{volume}{27}}, \bibinfo{pages}{243201}
  (\bibinfo{year}{2015}),
  \urlprefix\url{http://stacks.iop.org/0953-8984/27/i=24/a=243201}.

\bibitem[{\citenamefont{{Sato} and {Ando}}(2016)}]{sato_ando_review_arXiv_16}
\bibinfo{author}{\bibfnamefont{M.}~\bibnamefont{{Sato}}} \bibnamefont{and}
  \bibinfo{author}{\bibfnamefont{Y.}~\bibnamefont{{Ando}}},
  \bibinfo{journal}{ArXiv e-prints}  (\bibinfo{year}{2016}),
  \eprint{1608.03395}.

\bibitem[{\citenamefont{Read and Green}(2000)}]{read_green_PRB_00}
\bibinfo{author}{\bibfnamefont{N.}~\bibnamefont{Read}} \bibnamefont{and}
  \bibinfo{author}{\bibfnamefont{D.}~\bibnamefont{Green}},
  \bibinfo{journal}{Phys. Rev. B} \textbf{\bibinfo{volume}{61}},
  \bibinfo{pages}{10267} (\bibinfo{year}{2000}),
  \urlprefix\url{https://link.aps.org/doi/10.1103/PhysRevB.61.10267}.

\bibitem[{\citenamefont{Kallin and
  Berlinsky}(2016)}]{kallin_berlinsky_rep_prog_phys_16}
\bibinfo{author}{\bibfnamefont{C.}~\bibnamefont{Kallin}} \bibnamefont{and}
  \bibinfo{author}{\bibfnamefont{J.}~\bibnamefont{Berlinsky}},
  \bibinfo{journal}{Reports on Progress in Physics}
  \textbf{\bibinfo{volume}{79}}, \bibinfo{pages}{054502}
  (\bibinfo{year}{2016}),
  \urlprefix\url{http://stacks.iop.org/0034-4885/79/i=5/a=054502}.

\bibitem[{\citenamefont{Maeno et~al.}(2012)\citenamefont{Maeno, Kittaka,
  Nomura, Yonezawa, and Ishida}}]{maeno_review_12}
\bibinfo{author}{\bibfnamefont{Y.}~\bibnamefont{Maeno}},
  \bibinfo{author}{\bibfnamefont{S.}~\bibnamefont{Kittaka}},
  \bibinfo{author}{\bibfnamefont{T.}~\bibnamefont{Nomura}},
  \bibinfo{author}{\bibfnamefont{S.}~\bibnamefont{Yonezawa}}, \bibnamefont{and}
  \bibinfo{author}{\bibfnamefont{K.}~\bibnamefont{Ishida}},
  \bibinfo{journal}{Journal of the Physical Society of Japan}
  \textbf{\bibinfo{volume}{81}}, \bibinfo{pages}{011009}
  (\bibinfo{year}{2012}), \eprint{http://dx.doi.org/10.1143/JPSJ.81.011009},
  \urlprefix\url{http://dx.doi.org/10.1143/JPSJ.81.011009}.

\bibitem[{\citenamefont{Smidman et~al.}(2017)\citenamefont{Smidman, Salamon,
  Yuan, and Agterberg}}]{agterberg_review_17}
\bibinfo{author}{\bibfnamefont{M.}~\bibnamefont{Smidman}},
  \bibinfo{author}{\bibfnamefont{M.~B.} \bibnamefont{Salamon}},
  \bibinfo{author}{\bibfnamefont{H.~Q.} \bibnamefont{Yuan}}, \bibnamefont{and}
  \bibinfo{author}{\bibfnamefont{D.~F.} \bibnamefont{Agterberg}},
  \bibinfo{journal}{Reports on Progress in Physics}
  \textbf{\bibinfo{volume}{80}}, \bibinfo{pages}{036501}
  (\bibinfo{year}{2017}),
  \urlprefix\url{http://stacks.iop.org/0034-4885/80/i=3/a=036501}.

\bibitem[{\citenamefont{Fu and Kane}(2008)}]{fu_kane_08}
\bibinfo{author}{\bibfnamefont{L.}~\bibnamefont{Fu}} \bibnamefont{and}
  \bibinfo{author}{\bibfnamefont{C.~L.} \bibnamefont{Kane}},
  \bibinfo{journal}{Phys. Rev. Lett.} \textbf{\bibinfo{volume}{100}},
  \bibinfo{pages}{096407} (\bibinfo{year}{2008}),
  \urlprefix\url{https://link.aps.org/doi/10.1103/PhysRevLett.100.096407}.

\bibitem[{\citenamefont{Yanase and Sigrist}(2007)}]{yanase_sigrist_JPSJ_07}
\bibinfo{author}{\bibfnamefont{Y.}~\bibnamefont{Yanase}} \bibnamefont{and}
  \bibinfo{author}{\bibfnamefont{M.}~\bibnamefont{Sigrist}},
  \bibinfo{journal}{Journal of the Physical Society of Japan}
  \textbf{\bibinfo{volume}{76}}, \bibinfo{pages}{043712}
  (\bibinfo{year}{2007}),
  \urlprefix\url{http://dx.doi.org/10.1143/JPSJ.76.043712}.

\bibitem[{\citenamefont{Yanase and Sigrist}(2008)}]{yanase_sigrist_JPSJ_07b}
\bibinfo{author}{\bibfnamefont{Y.}~\bibnamefont{Yanase}} \bibnamefont{and}
  \bibinfo{author}{\bibfnamefont{M.}~\bibnamefont{Sigrist}},
  \bibinfo{journal}{Journal of the Physical Society of Japan}
  \textbf{\bibinfo{volume}{77}}, \bibinfo{pages}{124711}
  (\bibinfo{year}{2008}), \eprint{http://dx.doi.org/10.1143/JPSJ.77.124711},
  \urlprefix\url{http://dx.doi.org/10.1143/JPSJ.77.124711}.

\bibitem[{\citenamefont{Yokoyama et~al.}(2007)\citenamefont{Yokoyama, Onari,
  and Tanaka}}]{yokoyama_tanaka_PRB_07}
\bibinfo{author}{\bibfnamefont{T.}~\bibnamefont{Yokoyama}},
  \bibinfo{author}{\bibfnamefont{S.}~\bibnamefont{Onari}}, \bibnamefont{and}
  \bibinfo{author}{\bibfnamefont{Y.}~\bibnamefont{Tanaka}},
  \bibinfo{journal}{Phys. Rev. B} \textbf{\bibinfo{volume}{75}},
  \bibinfo{pages}{172511} (\bibinfo{year}{2007}),
  \urlprefix\url{http://link.aps.org/doi/10.1103/PhysRevB.75.172511}.

\bibitem[{\citenamefont{Hor et~al.}(2010)\citenamefont{Hor, Williams,
  Checkelsky, Roushan, Seo, Xu, Zandbergen, Yazdani, Ong, and
  Cava}}]{hor_ong_cava_PRL}
\bibinfo{author}{\bibfnamefont{Y.~S.} \bibnamefont{Hor}},
  \bibinfo{author}{\bibfnamefont{A.~J.} \bibnamefont{Williams}},
  \bibinfo{author}{\bibfnamefont{J.~G.} \bibnamefont{Checkelsky}},
  \bibinfo{author}{\bibfnamefont{P.}~\bibnamefont{Roushan}},
  \bibinfo{author}{\bibfnamefont{J.}~\bibnamefont{Seo}},
  \bibinfo{author}{\bibfnamefont{Q.}~\bibnamefont{Xu}},
  \bibinfo{author}{\bibfnamefont{H.~W.} \bibnamefont{Zandbergen}},
  \bibinfo{author}{\bibfnamefont{A.}~\bibnamefont{Yazdani}},
  \bibinfo{author}{\bibfnamefont{N.~P.} \bibnamefont{Ong}}, \bibnamefont{and}
  \bibinfo{author}{\bibfnamefont{R.~J.} \bibnamefont{Cava}},
  \bibinfo{journal}{Phys. Rev. Lett.} \textbf{\bibinfo{volume}{104}},
  \bibinfo{pages}{057001} (\bibinfo{year}{2010}),
  \urlprefix\url{https://link.aps.org/doi/10.1103/PhysRevLett.104.057001}.

\bibitem[{\citenamefont{Sasaki et~al.}(2011)\citenamefont{Sasaki, Kriener,
  Segawa, Yada, Tanaka, Sato, and Ando}}]{sasaki_ando_11}
\bibinfo{author}{\bibfnamefont{S.}~\bibnamefont{Sasaki}},
  \bibinfo{author}{\bibfnamefont{M.}~\bibnamefont{Kriener}},
  \bibinfo{author}{\bibfnamefont{K.}~\bibnamefont{Segawa}},
  \bibinfo{author}{\bibfnamefont{K.}~\bibnamefont{Yada}},
  \bibinfo{author}{\bibfnamefont{Y.}~\bibnamefont{Tanaka}},
  \bibinfo{author}{\bibfnamefont{M.}~\bibnamefont{Sato}}, \bibnamefont{and}
  \bibinfo{author}{\bibfnamefont{Y.}~\bibnamefont{Ando}},
  \bibinfo{journal}{Phys. Rev. Lett.} \textbf{\bibinfo{volume}{107}},
  \bibinfo{pages}{217001} (\bibinfo{year}{2011}),
  \urlprefix\url{https://link.aps.org/doi/10.1103/PhysRevLett.107.217001}.

\bibitem[{\citenamefont{Levy et~al.}(2013)\citenamefont{Levy, Zhang, Ha,
  Sharifi, Talin, Kuk, and Stroscio}}]{levy_stroscio_PRL_13}
\bibinfo{author}{\bibfnamefont{N.}~\bibnamefont{Levy}},
  \bibinfo{author}{\bibfnamefont{T.}~\bibnamefont{Zhang}},
  \bibinfo{author}{\bibfnamefont{J.}~\bibnamefont{Ha}},
  \bibinfo{author}{\bibfnamefont{F.}~\bibnamefont{Sharifi}},
  \bibinfo{author}{\bibfnamefont{A.~A.} \bibnamefont{Talin}},
  \bibinfo{author}{\bibfnamefont{Y.}~\bibnamefont{Kuk}}, \bibnamefont{and}
  \bibinfo{author}{\bibfnamefont{J.~A.} \bibnamefont{Stroscio}},
  \bibinfo{journal}{Phys. Rev. Lett.} \textbf{\bibinfo{volume}{110}},
  \bibinfo{pages}{117001} (\bibinfo{year}{2013}),
  \urlprefix\url{https://link.aps.org/doi/10.1103/PhysRevLett.110.117001}.

\bibitem[{\citenamefont{Brydon et~al.}(2014)\citenamefont{Brydon, Das~Sarma,
  Hui, and Sau}}]{brydon_sau_PRB_14}
\bibinfo{author}{\bibfnamefont{P.~M.~R.} \bibnamefont{Brydon}},
  \bibinfo{author}{\bibfnamefont{S.}~\bibnamefont{Das~Sarma}},
  \bibinfo{author}{\bibfnamefont{H.-Y.} \bibnamefont{Hui}}, \bibnamefont{and}
  \bibinfo{author}{\bibfnamefont{J.~D.} \bibnamefont{Sau}},
  \bibinfo{journal}{Phys. Rev. B} \textbf{\bibinfo{volume}{90}},
  \bibinfo{pages}{184512} (\bibinfo{year}{2014}),
  \urlprefix\url{https://link.aps.org/doi/10.1103/PhysRevB.90.184512}.

\bibitem[{\citenamefont{Chakhalian et~al.}(2007)\citenamefont{Chakhalian,
  Freeland, Habermeier, Cristiani, Khaliullin, van Veenendaal, and
  Keimer}}]{chakhalian_science_07}
\bibinfo{author}{\bibfnamefont{J.}~\bibnamefont{Chakhalian}},
  \bibinfo{author}{\bibfnamefont{J.~W.} \bibnamefont{Freeland}},
  \bibinfo{author}{\bibfnamefont{H.-U.} \bibnamefont{Habermeier}},
  \bibinfo{author}{\bibfnamefont{G.}~\bibnamefont{Cristiani}},
  \bibinfo{author}{\bibfnamefont{G.}~\bibnamefont{Khaliullin}},
  \bibinfo{author}{\bibfnamefont{M.}~\bibnamefont{van Veenendaal}},
  \bibnamefont{and} \bibinfo{author}{\bibfnamefont{B.}~\bibnamefont{Keimer}},
  \bibinfo{journal}{Science} \textbf{\bibinfo{volume}{318}},
  \bibinfo{pages}{1114} (\bibinfo{year}{2007}), ISSN \bibinfo{issn}{0036-8075},
  \eprint{http://science.sciencemag.org/content/318/5853/1114.full.pdf},
  \urlprefix\url{http://science.sciencemag.org/content/318/5853/1114}.

\bibitem[{\citenamefont{Reyren et~al.}(2007)\citenamefont{Reyren, Thiel,
  Caviglia, Kourkoutis, Hammerl, Richter, Schneider, Kopp, R{\"u}etschi,
  Jaccard et~al.}}]{reyren_mannhart_science_07}
\bibinfo{author}{\bibfnamefont{N.}~\bibnamefont{Reyren}},
  \bibinfo{author}{\bibfnamefont{S.}~\bibnamefont{Thiel}},
  \bibinfo{author}{\bibfnamefont{A.~D.} \bibnamefont{Caviglia}},
  \bibinfo{author}{\bibfnamefont{L.~F.} \bibnamefont{Kourkoutis}},
  \bibinfo{author}{\bibfnamefont{G.}~\bibnamefont{Hammerl}},
  \bibinfo{author}{\bibfnamefont{C.}~\bibnamefont{Richter}},
  \bibinfo{author}{\bibfnamefont{C.~W.} \bibnamefont{Schneider}},
  \bibinfo{author}{\bibfnamefont{T.}~\bibnamefont{Kopp}},
  \bibinfo{author}{\bibfnamefont{A.-S.} \bibnamefont{R{\"u}etschi}},
  \bibinfo{author}{\bibfnamefont{D.}~\bibnamefont{Jaccard}},
  \bibnamefont{et~al.}, \bibinfo{journal}{Science}
  \textbf{\bibinfo{volume}{317}}, \bibinfo{pages}{1196} (\bibinfo{year}{2007}),
  ISSN \bibinfo{issn}{0036-8075},
  \eprint{http://science.sciencemag.org/content/317/5842/1196.full.pdf},
  \urlprefix\url{http://science.sciencemag.org/content/317/5842/1196}.

\bibitem[{\citenamefont{Mizukami et~al.}(2011)\citenamefont{Mizukami, Shishido,
  Shibauchi, Shimozawa, Yasumoto, Watanabe, Yamashita, Ikeda, Terashima,
  Kontani et~al.}}]{mizukami_matsuda_nat_phys_11}
\bibinfo{author}{\bibfnamefont{Y.}~\bibnamefont{Mizukami}},
  \bibinfo{author}{\bibfnamefont{H.}~\bibnamefont{Shishido}},
  \bibinfo{author}{\bibfnamefont{T.}~\bibnamefont{Shibauchi}},
  \bibinfo{author}{\bibfnamefont{M.}~\bibnamefont{Shimozawa}},
  \bibinfo{author}{\bibfnamefont{S.}~\bibnamefont{Yasumoto}},
  \bibinfo{author}{\bibfnamefont{D.}~\bibnamefont{Watanabe}},
  \bibinfo{author}{\bibfnamefont{M.}~\bibnamefont{Yamashita}},
  \bibinfo{author}{\bibfnamefont{H.}~\bibnamefont{Ikeda}},
  \bibinfo{author}{\bibfnamefont{T.}~\bibnamefont{Terashima}},
  \bibinfo{author}{\bibfnamefont{H.}~\bibnamefont{Kontani}},
  \bibnamefont{et~al.}, \bibinfo{journal}{Nat Phys}
  \textbf{\bibinfo{volume}{7}}, \bibinfo{pages}{849} (\bibinfo{year}{2011}),
  \urlprefix\url{http://dx.doi.org/10.1038/nphys2112}.

\bibitem[{\citenamefont{Shimozawa et~al.}(2014)\citenamefont{Shimozawa, Goh,
  Endo, Kobayashi, Watashige, Mizukami, Ikeda, Shishido, Yanase, Terashima
  et~al.}}]{matsuda_PRL_14}
\bibinfo{author}{\bibfnamefont{M.}~\bibnamefont{Shimozawa}},
  \bibinfo{author}{\bibfnamefont{S.~K.} \bibnamefont{Goh}},
  \bibinfo{author}{\bibfnamefont{R.}~\bibnamefont{Endo}},
  \bibinfo{author}{\bibfnamefont{R.}~\bibnamefont{Kobayashi}},
  \bibinfo{author}{\bibfnamefont{T.}~\bibnamefont{Watashige}},
  \bibinfo{author}{\bibfnamefont{Y.}~\bibnamefont{Mizukami}},
  \bibinfo{author}{\bibfnamefont{H.}~\bibnamefont{Ikeda}},
  \bibinfo{author}{\bibfnamefont{H.}~\bibnamefont{Shishido}},
  \bibinfo{author}{\bibfnamefont{Y.}~\bibnamefont{Yanase}},
  \bibinfo{author}{\bibfnamefont{T.}~\bibnamefont{Terashima}},
  \bibnamefont{et~al.}, \bibinfo{journal}{Phys. Rev. Lett.}
  \textbf{\bibinfo{volume}{112}}, \bibinfo{pages}{156404}
  (\bibinfo{year}{2014}),
  \urlprefix\url{http://link.aps.org/doi/10.1103/PhysRevLett.112.156404}.

\bibitem[{\citenamefont{Caviglia et~al.}(2008)\citenamefont{Caviglia, Gariglio,
  Reyren, Jaccard, Schneider, Gabay, Thiel, Hammerl, Mannhart, and
  Triscone}}]{caviglia_triscone_Nature_08}
\bibinfo{author}{\bibfnamefont{A.~D.} \bibnamefont{Caviglia}},
  \bibinfo{author}{\bibfnamefont{S.}~\bibnamefont{Gariglio}},
  \bibinfo{author}{\bibfnamefont{N.}~\bibnamefont{Reyren}},
  \bibinfo{author}{\bibfnamefont{D.}~\bibnamefont{Jaccard}},
  \bibinfo{author}{\bibfnamefont{T.}~\bibnamefont{Schneider}},
  \bibinfo{author}{\bibfnamefont{M.}~\bibnamefont{Gabay}},
  \bibinfo{author}{\bibfnamefont{S.}~\bibnamefont{Thiel}},
  \bibinfo{author}{\bibfnamefont{G.}~\bibnamefont{Hammerl}},
  \bibinfo{author}{\bibfnamefont{J.}~\bibnamefont{Mannhart}}, \bibnamefont{and}
  \bibinfo{author}{\bibfnamefont{J.~M.} \bibnamefont{Triscone}},
  \bibinfo{journal}{Nature} \textbf{\bibinfo{volume}{456}},
  \bibinfo{pages}{624} (\bibinfo{year}{2008}),
  \urlprefix\url{http://dx.doi.org/10.1038/nature07576}.

\bibitem[{\citenamefont{Stock et~al.}(2008)\citenamefont{Stock, Broholm, Hudis,
  Kang, and Petrovic}}]{stock_magnetic_fluc_CeCoIn5}
\bibinfo{author}{\bibfnamefont{C.}~\bibnamefont{Stock}},
  \bibinfo{author}{\bibfnamefont{C.}~\bibnamefont{Broholm}},
  \bibinfo{author}{\bibfnamefont{J.}~\bibnamefont{Hudis}},
  \bibinfo{author}{\bibfnamefont{H.~J.} \bibnamefont{Kang}}, \bibnamefont{and}
  \bibinfo{author}{\bibfnamefont{C.}~\bibnamefont{Petrovic}},
  \bibinfo{journal}{Phys. Rev. Lett.} \textbf{\bibinfo{volume}{100}},
  \bibinfo{pages}{087001} (\bibinfo{year}{2008}),
  \urlprefix\url{https://link.aps.org/doi/10.1103/PhysRevLett.100.087001}.

\bibitem[{\citenamefont{R\o{}mer et~al.}(2015)\citenamefont{R\o{}mer, Kreisel,
  Eremin, Malakhov, Maier, Hirschfeld, and Andersen}}]{roemer_PRB_15}
\bibinfo{author}{\bibfnamefont{A.~T.} \bibnamefont{R\o{}mer}},
  \bibinfo{author}{\bibfnamefont{A.}~\bibnamefont{Kreisel}},
  \bibinfo{author}{\bibfnamefont{I.}~\bibnamefont{Eremin}},
  \bibinfo{author}{\bibfnamefont{M.~A.} \bibnamefont{Malakhov}},
  \bibinfo{author}{\bibfnamefont{T.~A.} \bibnamefont{Maier}},
  \bibinfo{author}{\bibfnamefont{P.~J.} \bibnamefont{Hirschfeld}},
  \bibnamefont{and} \bibinfo{author}{\bibfnamefont{B.~M.}
  \bibnamefont{Andersen}}, \bibinfo{journal}{Phys. Rev. B}
  \textbf{\bibinfo{volume}{92}}, \bibinfo{pages}{104505}
  (\bibinfo{year}{2015}),
  \urlprefix\url{http://link.aps.org/doi/10.1103/PhysRevB.92.104505}.

\bibitem[{\citenamefont{Scalapino et~al.}(1986)\citenamefont{Scalapino, Loh,
  and Hirsch}}]{scalapino_hirsch_86}
\bibinfo{author}{\bibfnamefont{D.~J.} \bibnamefont{Scalapino}},
  \bibinfo{author}{\bibfnamefont{E.}~\bibnamefont{Loh}}, \bibnamefont{and}
  \bibinfo{author}{\bibfnamefont{J.~E.} \bibnamefont{Hirsch}},
  \bibinfo{journal}{Phys. Rev. B} \textbf{\bibinfo{volume}{34}},
  \bibinfo{pages}{8190} (\bibinfo{year}{1986}),
  \urlprefix\url{https://link.aps.org/doi/10.1103/PhysRevB.34.8190}.

\bibitem[{\citenamefont{Hlubina}(1999)}]{hlubina_PRB_99}
\bibinfo{author}{\bibfnamefont{R.}~\bibnamefont{Hlubina}},
  \bibinfo{journal}{Phys. Rev. B} \textbf{\bibinfo{volume}{59}},
  \bibinfo{pages}{9600} (\bibinfo{year}{1999}),
  \urlprefix\url{https://link.aps.org/doi/10.1103/PhysRevB.59.9600}.

\bibitem[{\citenamefont{Meza and Riera}(2014)}]{riera_PRB_14}
\bibinfo{author}{\bibfnamefont{G.~A.} \bibnamefont{Meza}} \bibnamefont{and}
  \bibinfo{author}{\bibfnamefont{J.~A.} \bibnamefont{Riera}},
  \bibinfo{journal}{Phys. Rev. B} \textbf{\bibinfo{volume}{90}},
  \bibinfo{pages}{085107} (\bibinfo{year}{2014}),
  \urlprefix\url{https://link.aps.org/doi/10.1103/PhysRevB.90.085107}.

\bibitem[{sup()}]{supp_info}
\bibinfo{note}{See Supplemental Material for a derivation of Eqs. (4), (5), and
  (6) and additional plots of the spin susceptibility and the pairing
  strengths}.

\bibitem[{foo({\natexlab{a}})}]{footnoteWeakCoupling}
\bibinfo{note}{Since we focus on weak coupling, it is sufficient to consider
  only the static $\omega=0$ pairing interaction.}

\bibitem[{\citenamefont{Frigeri et~al.}(2006)\citenamefont{Frigeri, Agterberg,
  Milat, and Sigrist}}]{frigeri_sigrist_EPJB_2006}
\bibinfo{author}{\bibfnamefont{P.~A.} \bibnamefont{Frigeri}},
  \bibinfo{author}{\bibfnamefont{D.~F.} \bibnamefont{Agterberg}},
  \bibinfo{author}{\bibfnamefont{I.}~\bibnamefont{Milat}}, \bibnamefont{and}
  \bibinfo{author}{\bibfnamefont{M.}~\bibnamefont{Sigrist}},
  \bibinfo{journal}{The European Physical Journal B - Condensed Matter and
  Complex Systems} \textbf{\bibinfo{volume}{54}}, \bibinfo{pages}{435}
  (\bibinfo{year}{2006}), ISSN \bibinfo{issn}{1434-6036},
  \urlprefix\url{http://dx.doi.org/10.1140/epjb/e2007-00019-5}.

\bibitem[{\citenamefont{Dolgov and Golubov}(2008)}]{dolgov_golubov_PRB_08}
\bibinfo{author}{\bibfnamefont{O.~V.} \bibnamefont{Dolgov}} \bibnamefont{and}
  \bibinfo{author}{\bibfnamefont{A.~A.} \bibnamefont{Golubov}},
  \bibinfo{journal}{Phys. Rev. B} \textbf{\bibinfo{volume}{77}},
  \bibinfo{pages}{214526} (\bibinfo{year}{2008}),
  \urlprefix\url{https://link.aps.org/doi/10.1103/PhysRevB.77.214526}.

\bibitem[{\citenamefont{Dolgov et~al.}(2009)\citenamefont{Dolgov, Mazin,
  Parker, and Golubov}}]{dolgov_mazin_PRB_09}
\bibinfo{author}{\bibfnamefont{O.~V.} \bibnamefont{Dolgov}},
  \bibinfo{author}{\bibfnamefont{I.~I.} \bibnamefont{Mazin}},
  \bibinfo{author}{\bibfnamefont{D.}~\bibnamefont{Parker}}, \bibnamefont{and}
  \bibinfo{author}{\bibfnamefont{A.~A.} \bibnamefont{Golubov}},
  \bibinfo{journal}{Phys. Rev. B} \textbf{\bibinfo{volume}{79}},
  \bibinfo{pages}{060502} (\bibinfo{year}{2009}),
  \urlprefix\url{https://link.aps.org/doi/10.1103/PhysRevB.79.060502}.

\bibitem[{\citenamefont{Hirschfeld et~al.}(2011)\citenamefont{Hirschfeld,
  Korshunov, and Mazin}}]{hirschfeld_mazin_rep_prog_11}
\bibinfo{author}{\bibfnamefont{P.~J.} \bibnamefont{Hirschfeld}},
  \bibinfo{author}{\bibfnamefont{M.~M.} \bibnamefont{Korshunov}},
  \bibnamefont{and} \bibinfo{author}{\bibfnamefont{I.~I.} \bibnamefont{Mazin}},
  \bibinfo{journal}{Reports on Progress in Physics}
  \textbf{\bibinfo{volume}{74}}, \bibinfo{pages}{124508}
  (\bibinfo{year}{2011}),
  \urlprefix\url{http://stacks.iop.org/0034-4885/74/i=12/a=124508}.

\bibitem[{foo({\natexlab{b}})}]{footnote_RPA_at_van_Hove}
\bibinfo{note}{{\clb The weak coupling approach neglects retardations and
  restricts the momenta to the Fermi surfaces. This makes the calculation of
  $\lambda_i$, Eq.~\eqref{pairing_strength}, overly sensitive to the density of
  states, which is large very close to the van Hove fillings. For this reason
  the weak coupling approach should be considered with care close to
  $n_{vH_i}$}}.

\bibitem[{\citenamefont{Frigeri et~al.}(2004)\citenamefont{Frigeri, Agterberg,
  Koga, and Sigrist}}]{frigeri_sigrist_PRL_04}
\bibinfo{author}{\bibfnamefont{P.~A.} \bibnamefont{Frigeri}},
  \bibinfo{author}{\bibfnamefont{D.~F.} \bibnamefont{Agterberg}},
  \bibinfo{author}{\bibfnamefont{A.}~\bibnamefont{Koga}}, \bibnamefont{and}
  \bibinfo{author}{\bibfnamefont{M.}~\bibnamefont{Sigrist}},
  \bibinfo{journal}{Phys. Rev. Lett.} \textbf{\bibinfo{volume}{92}},
  \bibinfo{pages}{097001} (\bibinfo{year}{2004}),
  \urlprefix\url{https://link.aps.org/doi/10.1103/PhysRevLett.92.097001}.

\bibitem[{\citenamefont{Yanase et~al.}(2003)\citenamefont{Yanase, Jujo, Nomura,
  Ikeda, Hotta, and Yamada}}]{YANASE20031}
\bibinfo{author}{\bibfnamefont{Y.}~\bibnamefont{Yanase}},
  \bibinfo{author}{\bibfnamefont{T.}~\bibnamefont{Jujo}},
  \bibinfo{author}{\bibfnamefont{T.}~\bibnamefont{Nomura}},
  \bibinfo{author}{\bibfnamefont{H.}~\bibnamefont{Ikeda}},
  \bibinfo{author}{\bibfnamefont{T.}~\bibnamefont{Hotta}}, \bibnamefont{and}
  \bibinfo{author}{\bibfnamefont{K.}~\bibnamefont{Yamada}},
  \bibinfo{journal}{Physics Reports} \textbf{\bibinfo{volume}{387}},
  \bibinfo{pages}{1 } (\bibinfo{year}{2003}), ISSN \bibinfo{issn}{0370-1573},
  \urlprefix\url{http://www.sciencedirect.com/science/article/pii/S0370157303003235}.

\bibitem[{\citenamefont{Metzner et~al.}(2012)\citenamefont{Metzner, Salmhofer,
  Honerkamp, Meden, and Sch\"onhammer}}]{fRG_review_metzner_RMP_12}
\bibinfo{author}{\bibfnamefont{W.}~\bibnamefont{Metzner}},
  \bibinfo{author}{\bibfnamefont{M.}~\bibnamefont{Salmhofer}},
  \bibinfo{author}{\bibfnamefont{C.}~\bibnamefont{Honerkamp}},
  \bibinfo{author}{\bibfnamefont{V.}~\bibnamefont{Meden}}, \bibnamefont{and}
  \bibinfo{author}{\bibfnamefont{K.}~\bibnamefont{Sch\"onhammer}},
  \bibinfo{journal}{Rev. Mod. Phys.} \textbf{\bibinfo{volume}{84}},
  \bibinfo{pages}{299} (\bibinfo{year}{2012}),
  \urlprefix\url{https://link.aps.org/doi/10.1103/RevModPhys.84.299}.

\bibitem[{foo({\natexlab{c}})}]{footnote_disorder}
\bibinfo{note}{In Ref.~\cite{karen_fu_disorder_SC_PRL_12} it was shown that
  spin-orbit locking protects three-dimensional odd-parity superconductors
  against disorder, due to an emergent chiral symmetry. A similar protection
  mechanism is expected to occur in the Rashba-Hubbard model. A detailed study
  of this is left for further investigations.}

\bibitem[{\citenamefont{Michaeli and Fu}(2012)}]{karen_fu_disorder_SC_PRL_12}
\bibinfo{author}{\bibfnamefont{K.}~\bibnamefont{Michaeli}} \bibnamefont{and}
  \bibinfo{author}{\bibfnamefont{L.}~\bibnamefont{Fu}}, \bibinfo{journal}{Phys.
  Rev. Lett.} \textbf{\bibinfo{volume}{109}}, \bibinfo{pages}{187003}
  (\bibinfo{year}{2012}),
  \urlprefix\url{https://link.aps.org/doi/10.1103/PhysRevLett.109.187003}.

\end{thebibliography}

%%%%%%%%%%%%%%%%
%%%%%%%%%%%%%%%%
%%%%%%%%%%%%%%%%
%%%%%%%%%%%%%%%%

 \clearpage

 \setcounter{equation}{0}
\setcounter{figure}{0}
\setcounter{table}{0}
\setcounter{page}{1}
\renewcommand{\theequation}{S\arabic{equation}}
\renewcommand{\thefigure}{S\arabic{figure}}
%\renewcommand{\bibnumfmt}[1]{[S#1]}
%\renewcommand{\citenumfont}[1]{S#1}
%%%%%%%%%%

%\onecolumn
\onecolumngrid
\begin{center}
\textbf{\large Supplemental material: \\
Mechanism for unconventional superconductivity in the hole-doped \\ Rashba-Hubbard model} \\
\medskip 

Andr\'es Greco$^1$ and Andreas P. Schnyder$^2$ \\
\smallskip

${}^1${\em Facultad de Ciencias Exactas, Ingenier\'{\i}a y Agrimensura and
Instituto de 
\\
F\'{\i}sica Rosario (UNR-CONICET), Av. Pellegrini 250, 2000 Rosario, Argentina}

${}^2${\em Max-Planck-Institute for Solid State Research, D-70569 Stuttgart, Germany}

\end{center}
 
\bigskip

\twocolumngrid
%%%%%%%%%%%%%%%%%%%%%%%%%%%%%%%%%%%
%%%%%%%%%%%%%%%%%%%%%%%%%%%%%%%%%%%

In this supplemental material we give details of the derivation of the RPA spin susceptibility and the effective pairing interactions. 
We also present additional plots of the spin susceptibility and the pairing strengths
$\lambda_i^{\textrm{eff}}$.

\section{I.~~~Derivation of the RPA spin susceptibility} 

%%%%%%%%%%%%%%%%%%%%%%%%%%%%%%
\begin{figure}[b!]
\centering
\includegraphics[width=\columnwidth,angle=-0]{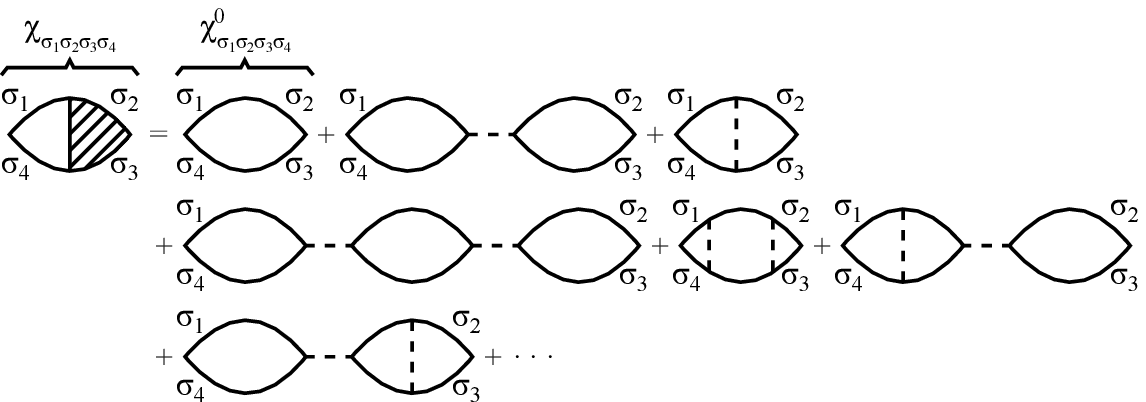}
\caption{  \label{infinity_sum}
The spin susceptibility $\chi_{\sigma_1 \sigma_2 \sigma_3 \sigma_4}$
is calculated within RPA as a sum of bubbles and ladders. In the diagrams only   
external indices ($\sigma_1,\sigma_2,\sigma_3$, and $\sigma_4$) are indicated. Internal indices or omitted and are assumed 
to be summed over. Note that due to finite SOC the contributions from bubble and ladder diagrams are mixed,
which is in contrast to the case without  SOC~\cite{roemer_PRB_15,scalapino_hirsch_86}.
} 
\end{figure}
%%%%%%%%%%%%%%%%%%%%%%%%%%%%%%

\begin{figure}[b!]
\centering
\includegraphics[width=5.cm,angle=-0]{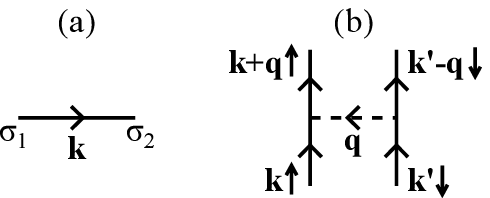}
\caption{  \label{propagators}
(a) Diagram of the  $2 \times 2$ fermionic propagator, Eq.~\eqref{2x2Green}. (b) Diagram of the four leg interaction vertex $U$, 
corresponding to the second term in Eq.~(1) of the main text. 
}
\end{figure}

Within the RPA the spin susceptibility $\chi_{\sigma_1 \sigma_2 \sigma_3 \sigma_4} ({\bf q},i\omega_l)$ 
in the spin basis
is obtained from an infinite sum of bubble and ladder diagrams, as shown in Fig.~\ref{infinity_sum}.
The solid lines in the diagrams correspond to the bare $2 \times 2$ 
fermionc propagator $G^{(0)}_{\sigma_1 \sigma_2} ( {\bf k} , i \nu_n )$ 
[see Eq.~\eqref{2x2Green} and Fig.~\ref{propagators}(a)]
and the dashed lines are the bare vertex $U$ [see Fig.~\ref{propagators} (b)]. 
The bare susceptibility, represented by a single bubble in Fig.~\ref{infinity_sum}, is given by
\begin{equation}
\chi^{(0)}_{\sigma_1 \sigma_2 \sigma_3 \sigma_4} ({\bf q},i\omega_l)=
\sum_{ {\bf k}, i\nu_n} G^{(0)}_{\sigma_1 \sigma_2}  ({\bf k},i\nu_n) 
G^{(0)}_{\sigma_3 \sigma_4}  ( {\bf k}+ {\bf q},i\nu_n+i\omega_l) ,
\end{equation}
with $\omega_l$ the bosonic Matsubara frequency, and 
\begin{equation}
G^{(0)}_{\sigma_1 \sigma_2} ( {\bf k} , i \nu_n ) = \left(\left[ i \nu_n  \sigma_0 - \hat{h} ({\bf k} ) \right]^{-1} \right)_{\sigma_1 \sigma_2}
\label{2x2Green}
\end{equation}
the  $2 \times 2$ bare Greens function 
with $ \nu_n = (2n+1) \pi / \beta$   the fermionic Matsubara frequency.
Arranging the sixteen matrix elements of $\chi^{(0)}_{\sigma_1 \sigma_2 \sigma_3 \sigma_4}$
 in a $4 \times 4$ matrix,
\begin{equation}
\hat{\chi}^{(0)}= 
\left(
\begin{array}{llll}
\chi^{(0)}_{\uparrow \uparrow \uparrow \uparrow}& \chi^{(0)}_{\uparrow \downarrow \uparrow \uparrow}& \chi^{(0)}_{\uparrow \uparrow \downarrow \uparrow}
&\chi^{(0)}_{\uparrow \downarrow \downarrow \uparrow}\\
\chi^{(0)}_{\uparrow \uparrow \uparrow \downarrow}& \chi^{(0)}_{\uparrow \downarrow \uparrow \downarrow}& \chi^{(0)}_{\uparrow \uparrow \downarrow \downarrow}
&\chi^{(0)}_{\uparrow \downarrow \downarrow \downarrow}\\
\chi^{(0)}_{\downarrow \uparrow \uparrow \uparrow}& \chi^{(0)}_{\downarrow \downarrow \uparrow \uparrow}& \chi^{(0)}_{\downarrow \uparrow \downarrow \uparrow}
&\chi^{(0)}_{\downarrow \downarrow \downarrow \uparrow}\\
\chi^{(0)}_{\downarrow \uparrow \uparrow \downarrow}& \chi^{(0)}_{\downarrow \downarrow \uparrow \downarrow}& \chi^{(0)}_{\downarrow \uparrow \downarrow \downarrow}
&\chi^{(0)}_{\downarrow \downarrow \downarrow \downarrow}
\end{array}
\right) \; ,
\end{equation}
the  infinite sum in  Fig.~\ref{infinity_sum} can be expressed analytically as~\cite{yanase_sigrist_JPSJ_07b}  
\begin{equation} \label{supp_RPA_suscep}
\hat{\chi} ({\bf q},i\omega_l)
=
\left[ I-   \hat{\chi}^{(0)} ( {\bf q}, i \omega_l )   \hat{U} \right]^{-1} 
\hat{\chi}^{(0)} ({\bf q},i\omega_l), 
\end{equation}
where  the dressed $4 \times 4$ matrix $\hat{\chi}$ contains the sixteen dressed susceptibilities arranged in similar form as in 
$\hat{\chi}^{(0)}$. 
The $4 \times 4$ interaction matrix $\hat{U}$ in Eq.~\eqref{supp_RPA_suscep} is off-diagonal with
\begin{equation}
\hat{U}= 
\left(
\begin{array}{llll}
0&0&0&U\\
0&0&-U&0\\
0&-U&0&0\\
U&0&0&0
\end{array}
\right) \; .
\end{equation}

%%%%%%%%%%%%%%%%%%%%%%%%%%%%%%
%%%%%%%%%%%%%%%%%%%%%%%%%%%%%%

The longitudinal and transversal susceptibilities can be computed 
in terms of the matrix elements $\chi_{\sigma_1 \sigma_2 \sigma_3 \sigma_4}$ as
\begin{equation}
\chi_{\textrm{long}}({\bf q},i\omega_l)=\chi_{\uparrow \uparrow \uparrow \uparrow}({\bf q},i\omega_l) - 
\chi_{\uparrow \downarrow \downarrow \uparrow} ({\bf q},i\omega_l)  
\end{equation}
and 
\begin{equation}
\chi_{\textrm{trans}}({\bf q},i\omega_l) = \chi_{\uparrow \uparrow \downarrow \downarrow}({\bf q},i\omega_l), 
\end{equation}
respectively. 
Note that the elements $(1,4)$ and $(4,1)$ of $\hat{U}$ enter in 
the bubble summation, while the elements $(2,3)$ and $(3,2)$ enter in the ladder summation in Fig.~\ref{infinity_sum}.

\section{II.~~~Derivation of the effective paring interactions} 
 
 %%%%%%%%%%%%%%%%%%%%%%%%%%%%%%
\begin{figure}[t!]
\centering
\includegraphics[width=5.cm,angle=-0]{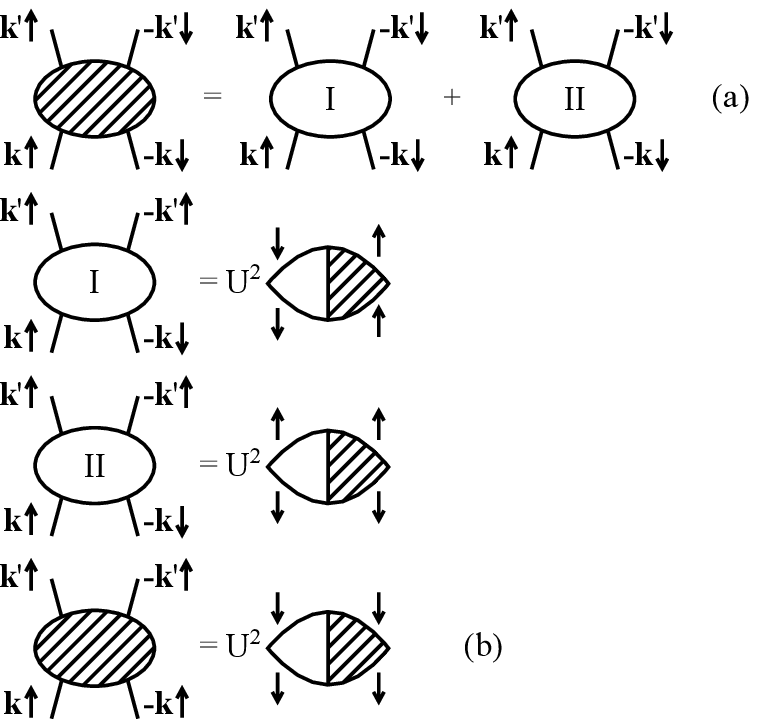}
\caption{  \label{effective_opp}
These diagrams depict the effective interactions
for (a) opposite spin projections and (b) same spin projections.
}
\end{figure}
%%%%%%%%%%%%%%%%%%%%%%%%%%%%%%

The effective spin-fluctuation-mediated interaction can be computed within the random 
phase approximation (RPA)\cite{roemer_PRB_15,scalapino_hirsch_86}.
It can be decomposed into same spin  projections $V^{\textrm{eff}}_{\textrm{same}} ({\bf k},{\bf k'})$ 
and opposite spin projections $V^{\textrm{eff}}_{\textrm{opp}}({\bf k}, {\bf k'})$. 
The effective interaction for opposite spin has two contributions, as shown in Fig.~\ref{effective_opp}(a),
which can be written as
\begin{equation}
V^{\textrm{eff}}_{\textrm{opp}}({\bf k}, {\bf k'})=U^2 \chi_{\sigma \bar{\sigma} \bar{\sigma} \sigma} ({\bf k-k'})+
U^2 \chi_{\sigma \sigma \bar{\sigma} \bar{\sigma }} ({\bf k+k'}) .
\end{equation}
The effective interaction for same spin projections, on the other hand,
has only  one contribution [see Fig.~\ref{effective_opp}(b)]
and is expressed as
\begin{equation}
V^{\textrm{eff}}_{\textrm{same}}({\bf k}, {\bf k'})=U^2 \chi_{\sigma \sigma \sigma \sigma} ({\bf k-k'}).
\end{equation}

Using these effective interactions and  employing a standard BCS approach, focusing on temperatures $T$ close to $T_\textrm{c}â$, 
we obtain the following linearized gap equations for the singlet ($\textrm{s}$) and 
triplet ($\textrm{t}$) channels in weak coupling
\begin{equation} \label{supp_linearized_gap_equations}
\Delta^{\textrm{s}/\textrm{t}}_{\alpha} (k ) 
= 
\ln \left( \frac{ 1.13  \, \omega_{\textrm{c}} }{   T_{\textrm{c}} } \right)
 \sum_\beta 
\int_{\textrm{FS}_\beta} 
\frac{ d k'}{ v^\beta_{\textrm{F}} ( k' ) } 
 V_{\textrm{s}/\textrm{t}}^{\textrm{eff}} (k, k') \Delta^{\textrm{s}/\textrm{t}}_{\beta} (k') ,
\end{equation}
where $\alpha, \beta = 1,2$ {\clb label the FS sheets}.
Here,  $k$ and $k'$ are restricted to the Fermi sheets  $\textrm{FS}_\alpha$ and 
 $\textrm{FS}_\beta$, respectively,   $v^{\alpha}_{\textrm{F}} ( {\bf k} )  = \left| \nabla_{\bf k} E^{\alpha}_{\bf k} \right|$
is the Fermi velocity, and
$\omega_{\textrm{c}}$ denotes the cutoff frequency, which
is given by the energy scale of the magnetic fluctuations.
In the case of singlet pairing the effective interaction $V_{\textrm{s}}^{\textrm{eff}}$ in Eq.~\eqref{supp_linearized_gap_equations} 
originates only from scattering between electrons with opposite spins, i.e., we set $V_{\textrm{s}}^{\textrm{eff}} (k, k') = V^{\textrm{eff}}_{\textrm{opp}}(  k ,   k' )$.
For triplet pairing, both equal- and opposite-spin scattering processes  
can yield a solution to Eq.~\eqref{supp_linearized_gap_equations}.
We therefore solve Eq.~\eqref{supp_linearized_gap_equations} for both $V_{\textrm{t}}^{\textrm{eff}}(k, k') = V^{\textrm{eff}}_{\textrm{same}}(  k ,  k' )$ and 
$V_{\textrm{t}}^{\textrm{eff}} (k, k') =  V^{\textrm{eff}}_{\textrm{opp}}(  k ,   k' )$.

%%%%%%%%%%%%%%%%%%%%%%%%%%%%%%
\begin{figure}[t!]
\centering
\includegraphics[width=\columnwidth,angle=-0]{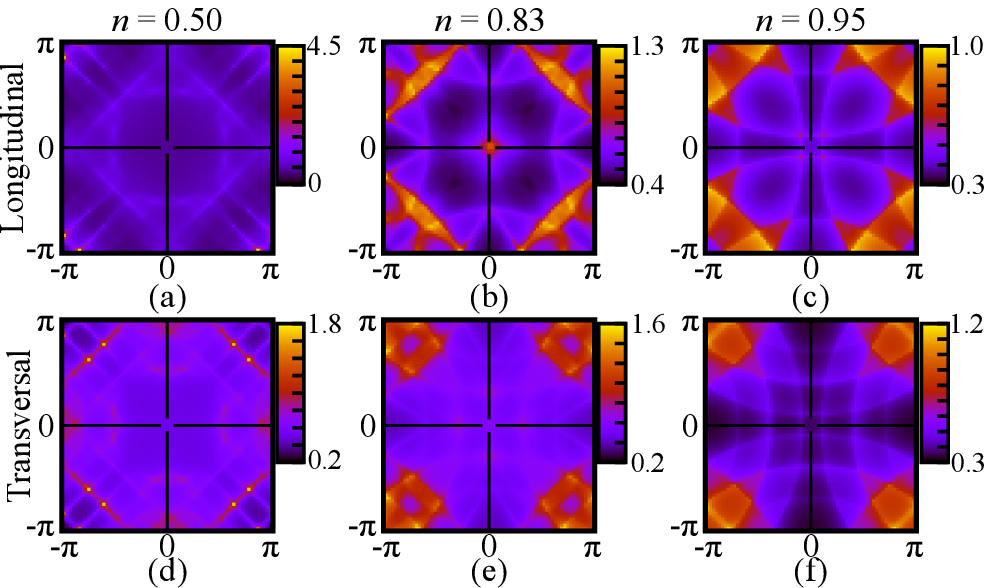}
\caption{  \label{suppFig_suscep_U22}
Longitudinal and transverse RPA spin susceptibilities 
for the fillings $n=0.50$, $n=0.83$, and $n=0.95$, respectively,
with $t^\prime = 0.3$, $V_{\textrm{so}}=0.5$,  $k_\textrm{B} T = 0.01$, and $U=2.2$.
The corresponding plots for $U=0.4$ are shown in Fig.~1 of the main text.
}
\end{figure}
%%%%%%%%%%%%%%%%%%%%%%%%%%%%%% 
 
In order to determine the symmetries of the pairing instabilities,   
we decompose the superconducting order parameter into an amplitude and a dimensionless symmetry function  
$\eta_i (k)$ for each pairing channel $i$.
That is, we write  $\Delta^{\textrm{s}/\textrm{t}}_\alpha ( k ) = \Delta_{\alpha} \eta_i (k) $, 
where  $\eta_i (k)$  describes the $k$ dependence
of each possible pairing symmetry on the square lattice:
\begin{eqnarray} \label{pairing_symmetries}
&&
\eta_{s} =1,
\;
\eta_{d_{x^2-y^2}}  = \cos k_x - \cos k_y,
\;
\eta_{d_{xy}}   = \sin k_x \sin k_y,
\nonumber\\
&&
\eta_p = \sin k_x , 
\; \textrm{and} \; \;
\eta_f = ( \cos k_x - \cos k_y) \sin k_x .
\end{eqnarray} 
Inserting the decomposition  $\Delta^{\textrm{s}/\textrm{t}}_\alpha ( k ) = \Delta_{\alpha} \eta_i (k) $ 
into Eq.~\eqref{supp_linearized_gap_equations},  multiplying both sides of the equation by $\eta_i(k)$, 
and integrating over  $k$ 
yields
\begin{equation}
\Delta_\alpha=\ln \left( \frac{ 1.13  \, \omega_{\textrm{c}} }{   T_{\textrm{c}} } \right) 
\sum_\beta \lambda_i^{\alpha \beta} \Delta_\beta  ,
\end{equation}
for each pairing channel $i$.
Here, $\lambda_i^{\alpha \beta}$ defines a  
$2 \times 2$ dimensionless pairing strength, 
which is given by~\cite{dolgov_golubov_PRB_08,dolgov_mazin_PRB_09,hirschfeld_mazin_rep_prog_11}
\begin{equation} \label{supp_pairing_strength}
\lambda_i^{\alpha \beta}
= 
- \frac{  
\mathlarger{\int}_{\textrm{FS}_\alpha} 
\frac{ d k}{ v^\alpha_{\textrm{F}} ( k ) }  
\mathlarger{\int}_{\textrm{FS}_\beta} 
\frac{ d k'}{ v^\beta_{\textrm{F}} ( k' ) } 
\eta_i (k) 
 V_{\textrm{s}/\textrm{t}}^{\textrm{eff}} (k, k')
\eta_i (k')
}
{
  2 \pi^2
\mathlarger{\int}_{\textrm{FS}_\beta} 
\frac{ d k^{\prime} }{ v^\beta_{F} (k^{\prime} ) }  
\left[ \eta_i(k^{\prime} ) \right]^2
}.
\end{equation}
The diagonal and off-diagonal elements of $\lambda_i^{\alpha \beta}$ represent
{\clb intra and inter FS pairing} strengths, respectively. 
The effective superconducting coupling constant $\lambda_i^{\textrm{eff}}$ for a given 
pairing channel $i$ is determined by the largest  eigenvalue
of the matrix $\lambda_i^{\alpha \beta}$~\cite{dolgov_golubov_PRB_08}, and allows to estimate $T_c$ in first approximation by
$T_c=1.13 \omega_c e^{-1/\lambda_i^{\textrm{eff}}}$.

\section{IV.~~~Additional Plots of the Spin Susceptibility}
%%%%%%%%%%%%%%%%%%%%%%%%%%%%%%
\begin{figure}[t!]
\centering
\includegraphics[width=\columnwidth,angle=-0]{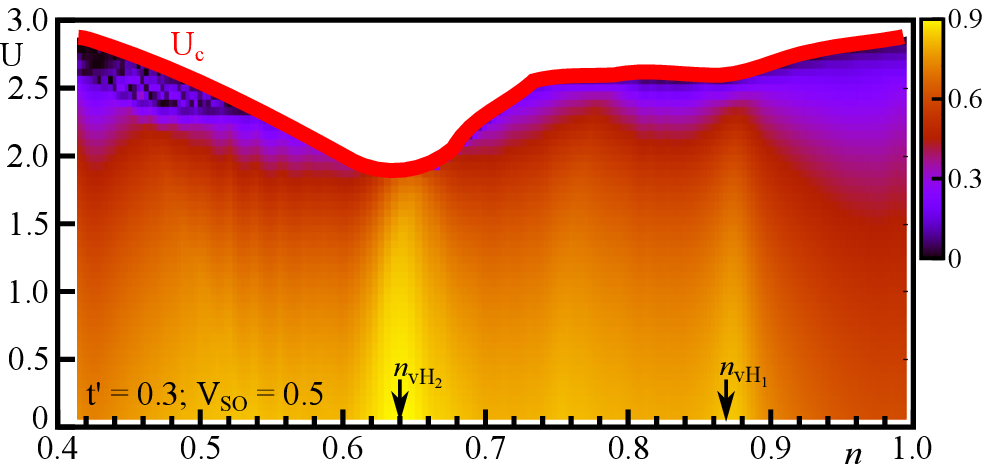}
\caption{  \label{suppFig_transverseSuscep}
Relative intensity of the ferromagnetic fluctuations in the transversal susceptibility 
as a function of interaction strength $U$ and filling $n$. The red line indicates the critical interaction strength $U_c$.
The corresponding plot for the longitudinal susceptibility is shown in Fig.~2 of the main text.
 }
\end{figure}
%%%%%%%%%%%%%%%%%%%%%%%%%%%%%%
In Fig.~\ref{suppFig_suscep_U22} we present plots of the longitudinal and transversal RPA spin susceptibility 
for three different fillings and $U=2.2$.
Similar to Fig.~1 in the main text, we observe strong ferromagnetic 
fluctuations for $n=0.83$ in the longitudinal susceptibility however, with lower intensity. 
Overall the ${\bf q}$ space structure of the spin fluctuations is quite similar to Fig.~1 in the main text. 
 
In Fig.~\ref{suppFig_transverseSuscep} we show
the relative strength of the ferromagnetic fluctuations in the transversal susceptibility as a function of $U$ and filling $n$.
In contrast to the longitudinal susceptibility shown in Fig.~2 in the main text, 
dominant ferromagnetic fluctuations only exist in very narrow regions around the two van Hove fillings
$n_{vH_1}$ and $n_{vH_2}$.\\

%%%%%%%%%%%%%%%%%%%%%%%%%%%%%%
\begin{figure}[t!]
\centering
\includegraphics[width=7.cm,angle=-0]{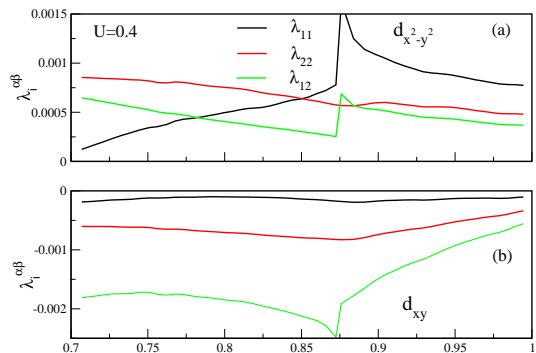}
\caption{  \label{mFig4}
Filling dependence of the {\clb intra and inter FS} pairing strengths for (a) the $d_{x^2-y^2}$-wave  and (b) the $d_{xy}$-wave pairing symmetry. 
The corresponding plot for the $f$-wave channel is shown in Fig.~3(c) of the main text. }
\end{figure}
%%%%%%%%%%%%%%%%%%%%%%%%%%%%%%

\section{V.~~~Additional Plots of the pairing strength}

In Figs.~\ref{mFig4}(a)-\ref{mFig4}(b) we present the filling dependence of
the {\clb intra FS} ($\lambda^{11}_i$ and $\lambda^{22}_i$) and the {\clb inter FS} ($\lambda^{12}_i$)
pairing strengths for the  $d_{x^2-y^2}$-wave and $d_{xy}$-wave pairing symmetry.
We observe that the $d_{x^2-y^2}$-wave instability is due to scattering within the  
Fermi sheet FS$_1$.  The case of $d_{xy}$-wave channel is interesting:
We find that  {\clb both inter and intra FS} pairing strengths  
are negative (repulsive), similar to the {\clb single FS} case ($V_{\textrm{so}}=0$) 
where $d_{xy}$-wave superconductivity is absent~\cite{roemer_PRB_15}.
However, in a two-band superconductor {\clb with two FSs}, as for $V_{\textrm{so}}\ne0$, large negative {\clb inter FS} interactions (green line) 
can drive pairing instabilites~\cite{dolgov_mazin_PRB_09,dolgov_golubov_PRB_08}.
In the present case this occurs for both the $d_{xy}$-wave channel [Fig.~\ref{mFig4}(b)] and the $p$-wave channel (not shown).

\end{document}